\documentstyle[epsfig]{mn}
\begin{document}
\input BoxedEPS.tex
\newcommand{\aip}{{\small ${\cal AIPS}$}}
\newcommand{\gtsim}{\mbox{{\raisebox{-0.4ex}{$\stackrel{>}{{\scriptstyle\sim}}
$}}}}
\newcommand{\ltsim}{\mbox{{\raisebox{-0.4ex}{$\stackrel{<}{{\scriptstyle\sim}}
$}}}}
\newcommand{\s}{$\stackrel{\rm s}{.}$}
\newcommand{\h}{$^{\rm h}$}
\newcommand{\m}{$^{\rm m}$}
\newcommand{\pp}{$\stackrel{\prime\prime}{.}$}
\newcommand{\de}{$^{\circ}$}
\newcommand{\p}{$^{\prime}$}
\newcommand{\arc}{$^{\prime\prime}$}
\newcommand{\marc}{^{\prime\prime}}
\newcommand{\rs}{{\em $r_s$}}
\newcommand{\DPM}{{\em DPM}}
\newcommand{\alf}{{\displaystyle\biggl({\nu_{\rm h} \over \nu_{\rm l}}\biggr)^{\alpha}} }

\newcommand{\figstart}[1]
    { \begin{figure}[htb]
      \begin{picture}(0,#1) }
\newcommand{\figend}[4]
    { \end{picture}
      \special{#1}
      \caption[#2]{#3}
      \label{#4}
      \end{figure} }
\newcommand{\fig}[5]
    { \figstart{#1}
      \figend{#2}{#3}{#4}{#5} }
\newcommand{\bHS}{\beta_{\mbox{\scriptsize HS}}}
\newcommand{\bBF}{\beta_{\mbox{\scriptsize BF}}}
\newcommand{\nT}{\nu_{\mbox{\scriptsize T}}}
\newcommand{\et}{E_{\mbox{\scriptsize T}}}
\newcommand{\nTn}{\nu_{\mbox{\scriptsize Tn}}}
\newcommand{\nTf}{\nu_{\mbox{\scriptsize Tf}}}
\newcommand{\tn}{\tau_{x\mbox{\scriptsize n}}}
\newcommand{\tf}{\tau_{x\mbox{\scriptsize f}}}
\newcommand{\xn}{x_{\mbox{\scriptsize n}}}
\newcommand{\xf}{x_{\mbox{\scriptsize f}}}
\newcommand{\yn}{y_{\mbox{\scriptsize n}}}
\newcommand{\yf}{y_{\mbox{\scriptsize f}}}
\newcommand{\lln}{l_{\mbox{\scriptsize n}}}
\newcommand{\llf}{l_{\mbox{\scriptsize f}}}
\newcommand{\Dn}{f(\Delta_{\mbox{\scriptsize n}})}
\newcommand{\Df}{f(\Delta_{\mbox{\scriptsize f}})}
\newcommand{\B}{\mbox{$B$}}
\newcommand{\Bo}{\mbox{$B$}_{0}}

\SetRokickiEPSFSpecial
\HideDisplacementBoxes


\title[ELAIS III:90$\mu m$ source counts]{The European Large
 Area {\em ISO} Survey  III: \\90$\mu m$  extragalactic source counts} 

\author[Andreas Efstathiou, Seb Oliver, Michael Rowan-Robinson,  et al]
{\parbox{159mm}{\begin{flushleft}
Andreas Efstathiou$^{1}$, Seb Oliver$^{1,2}$, Michael Rowan-Robinson$^{1}$,\\
{\LARGE C.~Surace$^{1}$,}
{\LARGE T.~Sumner$^{1}$,}
{\LARGE P.~H\'eraudeau$^{3}$,}
{\LARGE M.J.D.~Linden-V{\o}rnle$^{4}$,}
{\LARGE D.~Rigopoulou$^{5}$,}
{\LARGE S.~Serjeant$^{1}$,}
{\LARGE R.G.~Mann$^{1}$,}
{\LARGE C.J.~Cesarsky$^{6}$,}
{\LARGE L.~Danese$^{7}$,}
{\LARGE A.~Franceschini$^{8}$,}
{\LARGE R.~Genzel$^{5}$,}
{\LARGE A.~Lawrence$^{9}$,}
{\LARGE D.~Lemke$^{3}$,}
{\LARGE R.G.~McMahon$^{10}$,}
{\LARGE G.~Miley$^{11}$,}
{\LARGE J-L.~Puget$^{12}$,}
{\LARGE B.~Rocca-Volmerange$^{13}$,}
\end{flushleft}
}\vspace*{0.200cm}\\  
\parbox{159mm}{
$^{1}$ Astrophysics Group, Blackett Laboratory, Imperial College of 
Science Technology \& Medicine, Prince Consort
Rd., London.SW7 2BZ\\
$^{2}$ Astronomy Centre, University of Sussex, Falmer, Brighton BN1 9QJ\\
$^{3}$ Max-Planck-Institut f\"{u}r Astronomie, K\"{o}nigstuhl 17,
 D-69117, Heidelburg, Germany\\
$^{4}$Danish Space Research Institute, Juliane Maries Vej 30, DK--2100 Copenhagen
{\O}, Denmark\\
$^{5}$  Max-Planck-Institut f\"{u}r extraterrestrische Physik,
Postfach 1603, 85740 Garching, Germany\\
$^{6}$ CEA / SACLAY, 91191 Gif sur Yvette cedex, France\\
$^{7}$ SISSA, International School for Advanced Studies, 
Via Beirut 2-4, 34014 Trieste, Italy\\
$^{8}$ Dipartimento di Astronomia, Universita' di Padova,
 Vicolo Osservatorio 5, I-35122 Padova, Italy\\
$^{9}$ Institute for Astronomy, University of Edinburgh, Royal
Observatory, Blackford Hill, Edinburgh EH9 3HJ\\
$^{10}$ Institute of Astronomy, The Observatories, Madingley Road,
 Cambridge, CB3 0HA\\ 
$^{11}$ Leiden Observatory, P.O. Box 9513,  NL-2300 RA Leiden, The
Netherlands\\ 
$^{12}$ Institut d'Astrophysique Spatiale,  B\^{a}timent 121,
 Universit\'{e} Paris XI, 91405 Orsay cedex, France\\
$^{13}$ Institut d'Astrophysique de Paris, 98bis Boulevard Arago,
 F 75014 Paris, France\\ 
}}

\date{Submitted 18$^{th}$ October 1999; accepted 1$^{st}$ August 2000}
\maketitle
\begin{abstract}

 We present results and source counts at 90$\mu m$ extracted from the
 Preliminary Analysis of the European Large Area ISO Survey
 (ELAIS). The survey covered about 11.6deg$^2$ of the sky in four main
 areas and was carried out with the PHOT instrument onboard the
 Infrared Space Observatory (ISO).  The survey is at least an order of
 magnitude deeper than the IRAS 100$\mu m$ survey and is expected to
 provide constraints on the formation and evolution of galaxies. The
 majority of the detected sources are associated with galaxies on
 optical images. In some cases the optical associations are
 interacting pairs or small groups of galaxies suggesting the sample
 may include a significant fraction of luminous infrared galaxies. The
 source counts extracted from a reliable subset of the detected
 sources are in agreement with strongly evolving models of the
 starburst galaxy population.

\end{abstract}


\section{Introduction}

 As is well known, a large fraction of the bolometric luminosity of
 galaxies is emitted at far-infrared wavelengths as a result of
 reprocessing of starlight by dust.  The spectra of galaxies observed
 by IRAS typically peak in the 60-100$\mu m$ range. Observations of
 galaxies in the far-infrared are therefore essential for determining
 the star formation rate and dust masses of galaxies. Interest in
 dust-enshrouded galaxies has recently been renewed by the discovery
 of a far-infrared (140-850$\mu m$) background, thought to be due to
 discrete sources (Puget {\em et al.} 1996, Fixsen {\em et al.} 1998,
 Hauser {\em et al.}  1998, Lagache {\em et al.} 1999), and detections
 of high redshift galaxies with SCUBA (Smail {\em et al.} 1997, Hughes
 {\em et al.} 1998, Barger {\em et al.} 1998, Eales {\em et al.}
 1999), which are thought to be primeval analogues of the starburst
 galaxies observed locally.  Deep surveys in the far-infrared
 therefore promise to be very useful for understanding galaxy
 formation and evolution, as they can recover information missed by
 optical/UV studies (e.g. Steidel {\em et al.} 1996, Madau {\em et
 al.} 1996). Studies of infrared galaxies detected by IRAS suggest
 that violent episodes of star-formation are usually associated with
 very high dust optical depths (Condon {\em et al.} 1991,
 Rowan-Robinson \& Efstathiou 1993, Franceschini {\em et al.} 1994)
 which usually exceed unity even at 60$\mu m$. Models of dusty tori in
 AGN (Pier \& Krolik 1992, Granato \& Danese 1994, Efstathiou \&
 Rowan-Robinson 1995) also suggest that the optical depths of type 2
 objects are at least as high.  Observations of galaxies with ISO seem
 to have detected colder dust than observed by IRAS (Kr\"{u}gel {\em
 et al.} 1998, Bogun {\em et al.} 1996).

 The European Large Area ISO Survey (ELAIS; for a thorough description
 of the survey and its goals see Oliver {\em et al.}  2000, PAPER I),
 that surveyed about 12 square degrees of the sky at 15- and 90-$\mu
 m$ (and smaller areas at 6.7$\mu m$ and 175$\mu m$) in mainly four
 (low cirrus) areas spread over the northern and southern sky, and
 other ISO surveys (Kawara {\em et al.} 1998, Puget {\em et al.} 1999,
 Dole {\em et al.} 1999) represent the first opportunity since IRAS to
 study the properties and evolution of the far-infrared galaxy
 populations at intermediate redshift. An extensive follow-up
 programme from the radio to the X-rays, already under way, ensures
 that the ELAIS detected galaxies will be the subject of thorough
 study for the next few years.

 In this paper we describe the pipeline developed for the production
 of the ELAIS  Preliminary Analysis (EPA) 90$\mu m$ catalogue and present
 source counts. An accompanying paper by Serjeant {\em et al.} (2000; PAPER
 II) presents the 6.7 and 15$\mu m$ pipeline and counts.  Oliver {\em et al.}
 (in preparation) present an analysis of the multiply-covered
 ELAIS areas. 

 The paper is organized as follows: In section 2 we describe the data
 reduction and source extraction methods used. In section 3 we discuss
 tests for assessing the reliability and completeness of the survey
 including comparisons with a parallel pipeline to be described in a
 separate paper (Surace {\em et al.}  in preparation). In section 4 we present the
 source counts. In section 5 we discuss briefly our results and
 outline our conclusions.

\section{Observations and Reduction}

As described in detail in Paper I, the data consists of a
number of pointed observations (in raster mode) which use the 3x3
array of the ISOPHOT instrument (Lemke {\em et al.} 1996) onboard ISO
(Kessler {\em et al.} 1996). The area covered by each raster was typically
about a quarter of a square degree.  The integration time at each
raster position  varied but it was typically 20s. The initial
observing mode did not allow for any redundancy. Analysis
of some of the early rasters indicated that the efficiency and
reliability of the survey could be significantly improved with some
redundancy, so for the latter half or so of the survey we switched to
an `overlapping mode' so that each part of the sky observed thereafter
was covered at least twice. Some areas of the survey were observed on
more than one occasions thus providing further redundancy and depth.

The survey was carried out in four main areas (three in the northern and
one in the southern hemisphere) and some smaller areas of special
scientific interest (PAPER I).

The pipeline we developed for the Preliminary Analysis consists of two
basic stages. In the first stage we process the raw data using the
standard Phot Interactive Analysis (PIA) software (Gabriel {\em et al.} 1997)
until we derive a signal per 2s integration ramp. This involves
 discarding of  some readouts at the beginning of the ramp, and
treatment of cosmic rays.  The data stream includes data recorded
while the telescope is slewing between pointings (this typically takes
2 or 3 seconds per pointing).

The second stage of our analysis, performed with purpose-built IDL
routines, involves the extraction of point sources from the timeline
of each of the 9 pixels of the PHOT array. Given the poor
understanding of the instrument flat field at the time it was
considered premature to try to extract sources from maps.
Our source extraction method is therefore very different from that
employed in deep 175$\mu m$ surveys where multiple redundancy 
and  detector stability allow more traditional and reliable
techniques to be employed (Kawara {\em et al.} 1998, Puget {\em et al.} 1999,
 Dole {\em et al.} 1999).
 For each
pixel we ran an iterative median filter that removed outliers (arising
from cosmic rays) and determined the fractional deviation from the
background, $\rho $, for each ramp. Pointings for which the
weighted-average of $\rho$ over the pointing, $\bar{\rho}$, exceeded a
certain threshold (3$\sigma_{\bar{\rho}}$) were flagged as potential
point sources.

 The pipeline produced source lists on a raster-by-raster basis and
typically yielded $\sim 0.08$ detections per pointing per pixel. Unfortunately,
visual scanning of the time profiles of the potential sources revealed
that only a small fraction of them are likely to be real
sources. There are a number of effects that can give rise to spurious
sources and some of these are illustrated in Figure 1.

 Figure 1 shows the data stream of one pixel for about 6 pointings
 centred on a very clear detection of a source (at $t \approx 1050-1085s$)
 at different stages of the processing. Note the delay in the signal
 response to the change in illumination (about 2-3s) which is partly
 due to the movement of the telescope into position and partly due to
 the hysterisis of the detector. The latter is also responsible for the
 gradual fading of the signal at the end of the pointing. The
 occurence of these effects at the beginning and end of the pointing
 leaves little doubt as to  the reality of the detection.

 The source profiles are usually more complicated than the one shown
 in Figure 1, due to the occurence of cosmic ray events. Examples of
 such events can be seen on the same stream of data. Note for example
 the dips in the signal after the spikes at $t \approx 985s$ and
 $1115s$, and the fading profile after a spike at $t \approx
 1125s$. In addition, the C100 detectors sometimes display a drift
 behaviour which can give the appearance of a source if the peak of
 the cycle coincides with the centre of a pointing.

 While a
procedure for automatic filtering of false detections was developed it
was found to be unreliable. The time profiles of all potential sources
were therefore visually scanned by at least two observers in order to
construct the final source list. Five different classifications were
used ranging from  1 (definite detection; see Figure 1) to 5 (severely
affected by cosmic rays).

\begin{figure*}
\epsfig{file=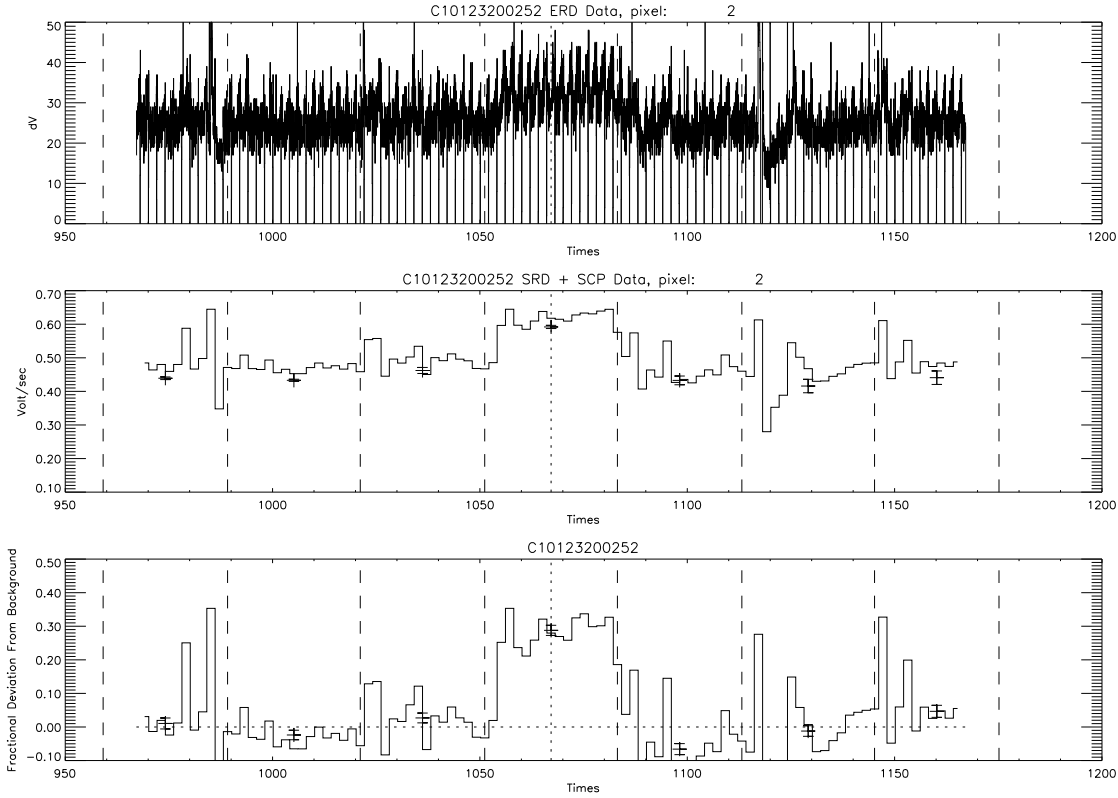,angle=0,width=15cm}
\caption{Timeline of a typical source with qual=1 (definite
detection). The top panel shows the raw data (in the form of the
difference $dV$ between successive readouts in volts) for about 100s centred
on the expected time of pointing at the source (dotted vertical
line). The dashed vertical lines show the time the telescope starts
slewing onto a new position. The middle panel shows the data at the
end of the PIA stage of the data reduction. The bottom panel shows the
fractional deviation from the background after the median filtering.}
\label{fig1}
\end{figure*}

\subsection{Flux Calibration}

The standard method of flux calibration of the PHOT instrument is to
make use of the internal Fine Calibration Source (FCS; Schulz {\em et
al.}  1998). The FCS has been calibrated using planets and asteroids at
bright fluxes and stars at  faint fluxes. The FCS is observed
immediately before and after each raster in order to monitor the
change in the responsivity of the detector over the course of the
raster. Despite the best efforts of the instrument team, the
calibration obtained in this way was considered until recently to be
of only qualitative value. It is now thought to be uncertain by about
20-30\% depending on the type of observation. This motivated us to
pursue an alternative method for the flux calibration for the EPA
 which makes use of celestial standards such as the 100$\mu m$
background measured by COBE/IRAS and IRAS sources.  The application of
a method based on celestial standards on a  large and
homogeneous dataset such as ELAIS can also provide an independent
check on the FCS method and provide insight into the characteristics
of the instrument.

 The basic steps followed in our method can be summarized as follows:
\begin{enumerate}

\item determine the flux per pixel $F_{\nu, det}$ (in Jy) of each
detection by multiplying $\bar{\rho} $ by the predicted
background flux incident on the pixel $F_\nu \equiv I_\nu \Omega$,
where $I_\nu$ is the predicted background intensity and $\Omega$ is
the solid angle ($0.44468 \times 10^{-7} sr$; Klaas et al 1994)
subtended by a C100 ($43.5'' \times 43.5''$) pixel.

\item find an empirical conversion factor $f_{p \rightarrow t}$ for
converting $F_{\nu, det}$ to a total flux for a point source centred
on the pixel $F_{\nu, src}$. This factor is determined by comparing
the corrected fluxes with those of calibration stars and IRAS sources
and includes corrections for the psf, straylight etc.  We will find
that $f_{p \rightarrow t}$ is the ratio of two factors $f_{bckg}$ and
$f_{psf}$.  The factor $f_{bckg}$ corrects our background estimate for
the effective solid angle, straylight etc. and $f_{psf}$ corrects for
the fact that only a fraction of the flux from a point source will be
recorded by a single-pixel detection.

\end{enumerate}

The great advantage of the method is its transparency. Its validity
ultimately depends on the accuracy with which the background can be
estimated (given that it is dominated by the zodiacal emission which
varies with the time of observation) and on the accuracy of the fluxes
of the sources used as calibrators.

In the rest of this section we describe how we estimate the
background, $f_{bckg}$ and $f_{psf}$, and test our calibration with
stars and IRAS sources. Readers who are not interested in the details
of these estimates and tests can find a summary of our results in
section 2.5.

\subsection{Background estimate}

 Initially,
we used the COBE background (obtained from the IRSKY service at
IPAC) which gave us very poor spatial resolution. Our approach was
subsequently refined to consider the separate contributions from the local
(zodiacal) background, which depends on the  time of
observation, and Galactic \& extragalactic components.

 The background is estimated by combining the contributions of Galactic
emission from the 100$\mu m$ maps of Schlegel {\em et al.} (1998) and a model
for the zodiacal light.

 The zodiacal light emission depends on the ecliptic latitude $\beta$ and the
solar elongation  $e$ (e.g. CAM manual). It is worth noting that due
to observing constraints for the satellite the solar elongation of most ISO
observations was in the range 60-120$^o$. With the exception of two
small areas (X3 and X6) $|\beta|$ for  the ELAIS areas
lies between 40 and 75$^o$.

 The zodiacal model used is based on the model described in the CAM
manual (page 77) which gives the zodiacal background at 10.9$\mu m$ as
a function of $\beta$ and (more crudely) $e$. The predicted 10.9$\mu
m$ background from this model is about 30\% higher than that measured
by ISOCAM towards the ecliptic plane ($e=104^o$ and $\beta=-2.4^o$;
Reach {\em et al.} 1996) with the quoted uncertainty of ISOCAM calibration at
the time being 15\%.

 To extrapolate to far-infrared wavelengths we assume that the
spectrum of the zodiacal light can be modelled as a blackbody at a
(constant) temperature of 275K (Hauser {\em et al.} 1984). This model gives a
ratio of 100$\mu m$/25$\mu m$ backgrounds of 0.16, in very good
agreement with the estimate of Schlegel {\em et al.} (1998). 

 In general, the temperature of the dust in the zodiacal cloud could
vary with $\beta $ and 
$e$ (as our line of sight through the zodiacal cloud samples dust at
varying distances from the sun) but this is unlikely to affect the
estimated background by more than 5-10\% (Schlegel {\em et al.}). The range of
$|\beta|$ and $e$ spanned by the ELAIS survey observations is also
quite narrow.

In Figure 2 we compare our background estimate $F_\nu$, with the flux
per pixel determined by PIA\footnote{We refer here to the quantity
phtiaap.mnfl produced by PIA Version 7.3} using the FCS calibration,
$F_\nu^{PIA}$. We find that the ratio of the two raster-averaged
estimates $f_{bckg} =<F_{\nu}^{PIA}> / <F_{\nu}>$ has a median value
of 1.7.  The origin of this difference is unclear but it may be
related to the incidence of straylight on the detector. What is very
encouraging is that if we multiply our estimate of the background by
$f_{bckg}$, it agrees with that determined by PIA within 15\%.

\begin{figure*}
\epsfig{file=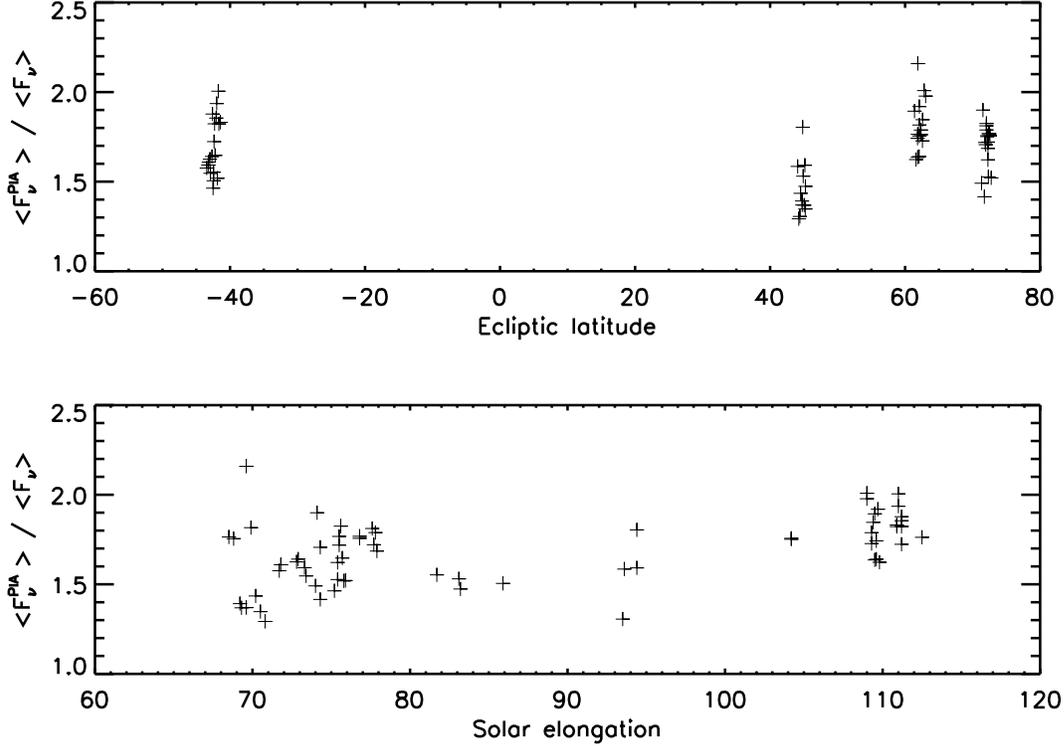,angle=0,width=15cm}
\caption{A plot of the ratio of the background obtained using PIA
(averaged over all pointings/pixels in a raster) $<F_\nu^{PIA}>$ and
that from our estimate from a model of the zodiacal background and
maps of galactic emission $<F_\nu>$ as a function of (a) ecliptic
latitude $\beta$ and (b) solar elongation.}
\label{fig2}
\end{figure*}

\subsection{ELAIS calibration stars}

 Given that there are no bright ($S > 0.6Jy$) 12$\mu m$ sources in the
 ELAIS fields (to avoid saturating ISOCAM) we don't expect any
 `photospheric' stars to be detected at 90$\mu m$ as these would have
 a $90\mu m$ flux $ \leq 10mJy$ . It is possible that there may be
 some stars with circumstellar dust shells, but the usefulness of such
 stars as calibrators would be limited.  In order to better determine
 the ELAIS calibration (as well as the general ISOPHOT calibration)
 three stars (HR6132, HR6464 and HR5981) close to the ELAIS fields
 were observed in mini-raster mode (a $3 \times 3$ raster with the
 star positioned at the centre of a different pixel in each
 pointing). These stars also formed part of the ISO ground based
 preparatory programme. Models for their far-IR spectra were
 constructed by fitting near-IR data and extrapolating to longer
 wavelengths as $\nu^2$ (Hammersley {\em et al.}  1998).  A more
 empirical approach was given by Cohen {\em et al.}  (1999). We use
 Hammersley's predictions for the two brighter stars but for HR5981 we
 use the prediction of Cohen as it is more reliable for cool
 stars. The predicted stellar fluxes (after convolving with the ISO
 90$\mu m$ filter response) lie in the range (68-311mJy) and so 
 extend the flux range to fainter fluxes than possible with the IRAS
 catalogued sources (see section 2.4). The IRAS fluxes of these stars
 agree very well with the model up to $60\mu m$. At $100\mu m$ results
 from SCANPI ({\em alias} ADDSCAN) for the two brightest stars, which
 show significant detections, also show an excess over the models by
 factors of 2-5. It is not clear whether this implies that the stars
 have an infrared-excess (e.g. due to dust) or whether the SCANPI
 results suffer from some artifact introduced in the processing (e.g
 by cirrus structure).  The faintest of the stars (HR5981) was
 observed twice on the same ISO orbit. The integration time per
 pointing in these mini-rasters (40s) is longer than that used for the
 bulk of the ELAIS survey. The mini-rasters for the calibration stars
 were processed in the same way as the survey rasters.

 In principle, each detection of each of the stars can be used to
 estimate $f_{p \rightarrow t}$, i.e. the correction factor for peak
 flux $F_{\nu, det}$ to total flux $F_{\nu, src}$ for our particular
 observing mode and in the ideal situation where a point source is
 centred in the pixel. We find the median value of $f_{p \rightarrow
 t}$ to be 2.8. In Figure 3 we plot the ratio of $f_{p \rightarrow t}
 F_{\nu, det}$ over the predicted stellar flux for each detection. As
 expected, the empirically estimated $f_{p \rightarrow t}$ is $\sim
 f_{bckg}/f_{psf}(0,0)$ where $f_{psf}(0,0)=0.6$ is the theoretically
 predicted fraction of total flux in the central pixel (Klaas {\em et
 al.} 1994, Laureijs 1999).

\begin{figure*}
\epsfig{file=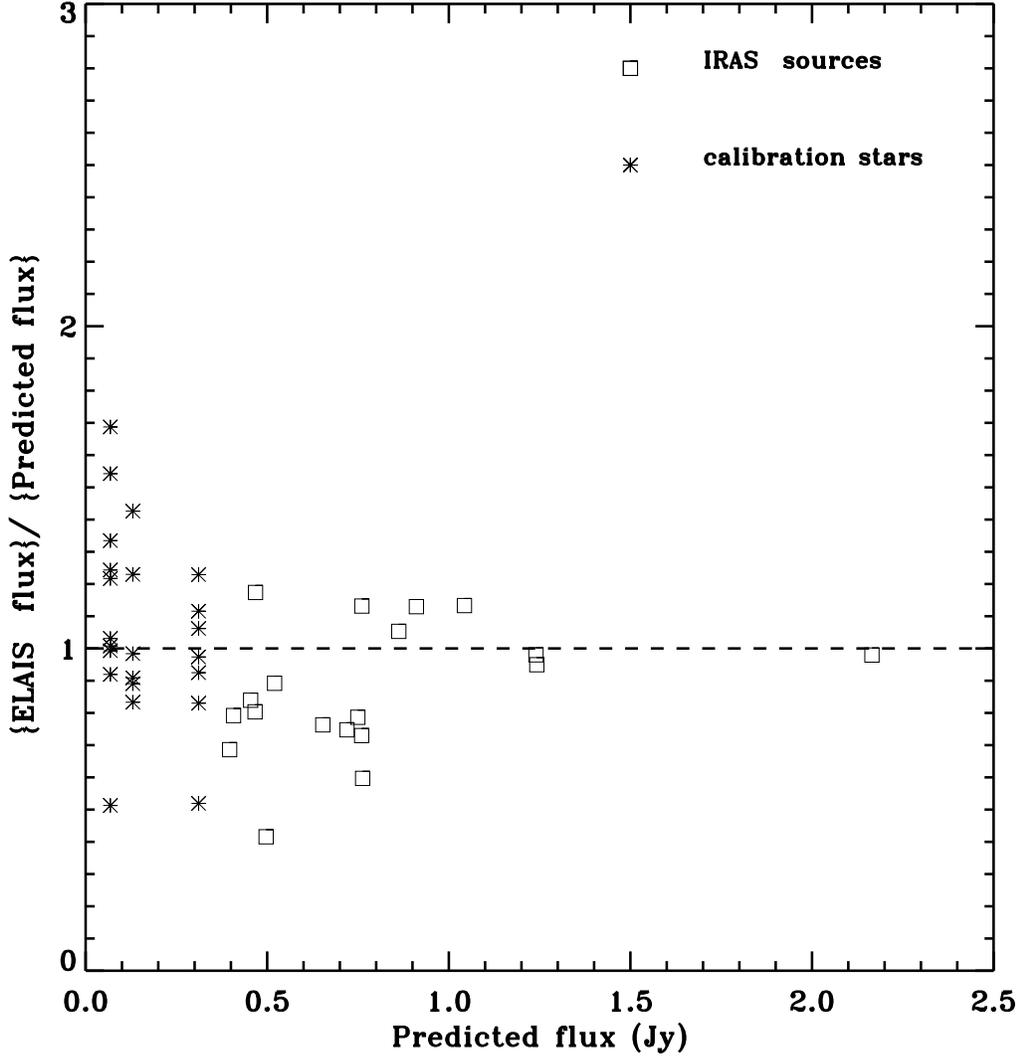,angle=90,width=15cm}
\caption{Comparison of ELAIS fluxes with IRAS fluxes (squares) and
theoretically predicted fluxes for the calibration stars (stars).  The
90$\mu m$ fluxes of the IRAS sources are estimated by linearly
interpolating between the 60 and 100$\mu m$ fluxes. The ratio of 90 to
100$\mu m$ flux is typically 0.8.  For the three calibration stars, we
plot all the detections (except those in the first and last pointings
which make the background estimation rather difficult). The scatter in
the ratio gives a measure of the reproducibility of the results. The
predicted fluxes of the stars are obtained by convolving the models of
Hammersley {\em et al.} (1998), for HR6132, HR6464, and Cohen {\em et
al.} (1999), for the fainter star HR5981, with the ISO filter response
function.}
\label{fig3}
\end{figure*}

\subsection{Comparison with IRAS sources}

While the ELAIS fields were chosen to avoid strong infrared sources,
there are a number of IRAS 100$\mu m$ sources in the PSC and FSC which
were detected in the survey, some of them in a number of pixels. The
fluxes of these sources lie in the range $0.5Jy \leq S(100) \leq 3Jy$
and all of them  also have 60$\mu m$ detections. Application of the
correction factor $f_{p \rightarrow t}$, derived from the calibration
observations, to the peak flux generally gave fluxes too low compared
with the IRAS fluxes (after correcting for the small change in
effective central wavelength of the IRAS and ISO filters) by factors
of 2-3. There are a number of possible reasons for this discrepancy:

\begin{enumerate}

\item Firstly, the optical identifications of these sources appear to
be galaxy pairs or galaxies with sizes comparable to or larger than
the pixel size (43.5''). The ISO telescope psf (for the 105$\mu m$
filter) has been approximately fitted with a gaussian with FWHM of
$50\arcsec\pm 5\arcsec$ (H\'eraudeau {\em et al.} 1997, MPIA internal
report). Clearly for such extended sources it is not appropriate to scale the
peak flux by $f_{p \rightarrow t}$.

\item As it was demonstrated by Hacking \& Houck (1987) in the Deep
IRAS survey in the North Ecliptic Pole, the 100$\mu m$ fluxes of the
survey proper become unreliable as the flux limit of the survey is
approached ($\sim 0.5$Jy at 100$\mu m$). The fluxes tend to be
overestimated by factors of up to 2 as they are boosted by positive
noise fluctuations  which are superimposed on the true source flux.

\item It is very unlikely that the source (even if it were pointlike)
would be centred in the pixel having the peak flux. In these
situations $f_{p \rightarrow t}$ should be considered as a lower limit
to the scale factor.

\item Due to the relatively short integration time per pointing it is
possible that for bright sources the signal does not reach its stabilized
value, so the flux  will be underestimated.

\end{enumerate}

 In order to determine whether a source is extended we need to know
 what fraction of the flux from a point source would fall in an
 adjacent, or diagonally placed, pixel. These fractions were estimated
 by Laureijs (1999) for a point source centred in a pixel. His results
 show that $f_{psf}(0,46)/f_{psf}(0,0)=0.06$ and
 $f_{psf}(46,46)/f_{psf}(0,0)=0.02$ where $f_{psf}(x,y)$ is the fraction of
 the flux in the pixel centred at $x,y$ arcseconds from the point
 source. This assumes all the pixels have the same roll angle.  Note
 that although each C100 pixel is 43.5'' square, the separation of
 pixel centers is 46'' (Klaas {\em et al.} 1994).

 According to this analysis we can assume that adjacent detections
 with flux greater than $\sim$ 15\% of the peak flux correspond to
 extended sources or members of group of sources. There are, of
 course, situations where this criterion will fail, that is when a
 point source lies at the corner or edge of a pixel. We have simulated
 this situation and found that we are unlikely to overestimate the
 flux by more than 30\%.  Laureijs also calculated the fraction of the
 psf in the whole array and found it to be $\sim 0.8$ for the C90
 filter, which also sets a limit of $\sim$ 30\% to the degree by which
 we would overestimate the flux of a point source in this case.

\subsection{Summary of calibration method}

 In summary, our strategy for flux calibration is as follows: for
 sources with more than 1 reliable detections, the total flux is
 calculated by ${{f_{bckg}} \over {f_{psf}(0,0)}} \sum_i F_{\nu,
 det}^i$, where $F_{\nu, det}^i$ are the fluxes of non-overlapping
 detections above 15\% of the peak. The factor $f_{bckg}$ is estimated
 to be 1.7. The factor $f_{psf}(0,0)$ is assumed to be 0.6 (Klaas et
 al 1994).  For single detections we calculate the flux as $f_{bckg}
 F_{\nu, det} / <f_{psf}>$ where $<f_{psf}>$ is estimated from
 simulations of a point source with a FWHM of 50$\arcsec$ at different
 positions in the pixel to be $\sim 0.4$.

In Figure 3 we compare the predicted   90$\mu m$ fluxes of the IRAS
sources with the fluxes calibrated in this way.  In conclusion, our
approach to the flux calibration gives fluxes that agree with
those of IRAS sources and bright stars (which bracket the flux range
covered by ELAIS) within 30-40\%.

\section{Reliability and completeness}

\subsection{Comparison with an independent pipeline}

 We have pursued a programme of detailed comparisons of the IC
 detection lists with those compiled by a pipeline developed in
 parallel (but somewhat later, to take advantage of a better
 understanding of the instrument) by the group at the
 Max-Planck-Institut f\"{u}r Astronomie (MPIA) in Heidelberg. Results
 from that pipeline will be described in detail by Surace {\em et al.}
 (in preparation). The detections obtained by the MPIA pipeline were
 also eyeballed with the same classification scheme applied to the IC
 detections.

 The sample of sources on which the preliminary
 90$\mu m$ counts are based is a reliable subset of the EPA
 list which was constructed as follows:

\begin{enumerate}

\item  A list of detections  extracted by  either pipeline  and
which received the top two eyeballing classifications (1 or 2) 
by at least 2 observers was first constructed.

\item The  combined list was then trimmed to include only  
 detections which received one of the top two  classifications 
 by at least 3  observers (275 detections). 

\item The detections were then merged with a near neighbour algorithm
(search radius of 1') to produce a source list consisting of 163 sources. 
Only one of the sources with $S> 100mJy$ was detected by the MPIA
pipeline only; otherwise, the lists from the two pipelines are in
complete agreement above this limit. 

\end{enumerate}

In Figure 4 we compare the fluxes of the QLA reliable detections ({\em
not} sources) as determined by the method described here
(i.e. $f_{bckg} F_{\nu, det}$) with those derived from the MPIA
pipeline. While there is considerable scatter (comparable to the
scatter seen in the comparison of the backgrounds) it is rather
encouraging that above 50mJy the fluxes agree within 40\%. At the
bright end the MPIA fluxes may be underestimated because of
self-subtraction. Below 50mJy there seems to be a group of
detections where the fluxes derived by the two methods differ by
factors of 2-3.

\begin{figure*}
\epsfig{file=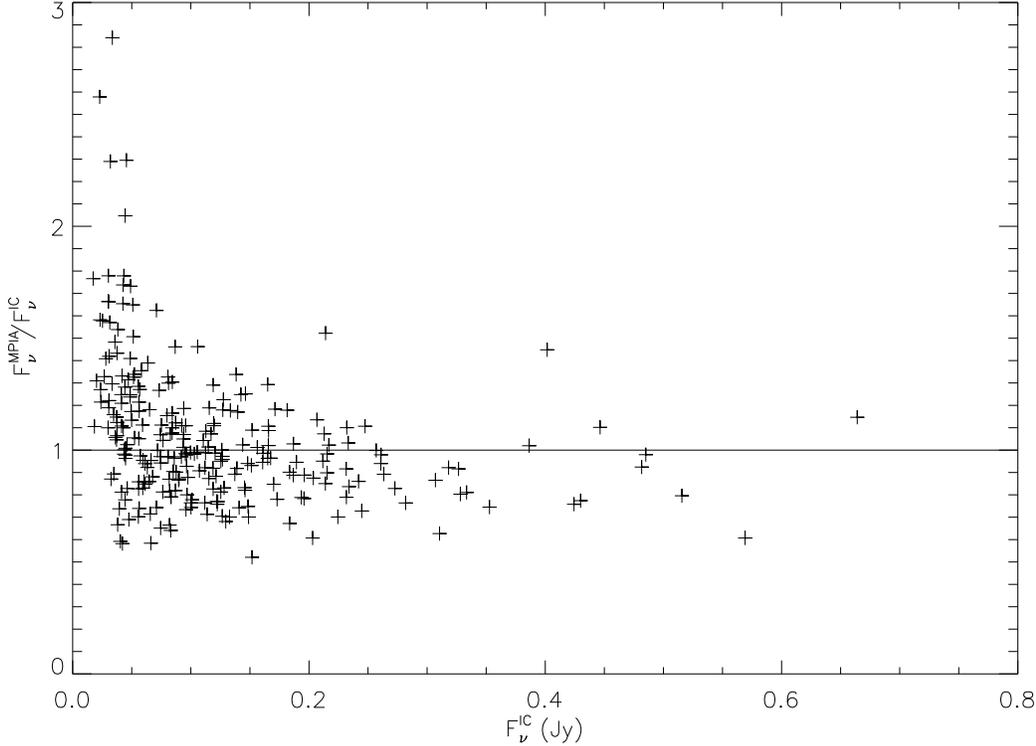,angle=0,width=15cm}
\caption{Comparison of the fluxes of the QLA reliable detections with those
derived from the MPIA pipeline.}
\label{fig4}
\end{figure*}


\subsection{Simulations}

 The data are strongly affected by cosmic rays and other detector
 effects so the coverage is very inhomogeneous.  In order to extract
 the source counts we need to quantify the incompleteness as a
 function of flux and position in the survey area.  There are two main
 causes of incompleteness. Firstly, sources that lie at the edges or
 corners of pixels are likely to be missed. In areas where there is
 redundancy (either because the overlapping mode is used or because
 the area has been re-observed) the probability of missing sources
 decreases. Secondly, some genuine sources may have been rejected at
 the eyeballing stage because they coincide with cosmic rays etc.

 To make a rough estimate of the degree of incompleteness due to these
 (and possibly other) effects we have simulated a number of rasters
 and processed them in an identical fashion as the survey rasters. The
 simulated data are constructed by adding scans of randomly placed
 sources onto real data extracted from several rasters (where the
 rasters and pixels are chosen at random). The detected sources were
 merged in the same way as the real sources. For practical reasons the
 simulated detections were eyeballed by a single observer (AE). This
 could introduce a bias in the completeness correction as we require
 confirmation (i.e. eyeballing class 1 or 2) by at least 3 observers
 for a detection to appear in our final list of real sources. To
 quantify the effect of this we can use the eyeballing experience of
 the whole survey to find the number of detections confirmed by AE
 which finally appeared in the combined list (186 out of 286). For
 comparison, the number of detections {\em not} confirmed by AE which
 appeared in the combined list is 34. As the number of detections
 confirmed by AE only (286), and by at least 3 observers (186+34), 
 are comparable we conclude that there is not a significant bias
 introduced by using the eyeballing results of a single observer for
 the purpose of estimating the completeness. In order to simulate
 regions of multiple redundancy we simulated rasters with different
 steps in one of the raster dimensions. Two possibilities were
 simulated: half an array step, to simulate regions of single
 redundancy, and quarter array step, to simulate regions of triple
 redundancy. The sources are calibrated in the same way as the real
 sources.

\begin{figure*}
\epsfig{file=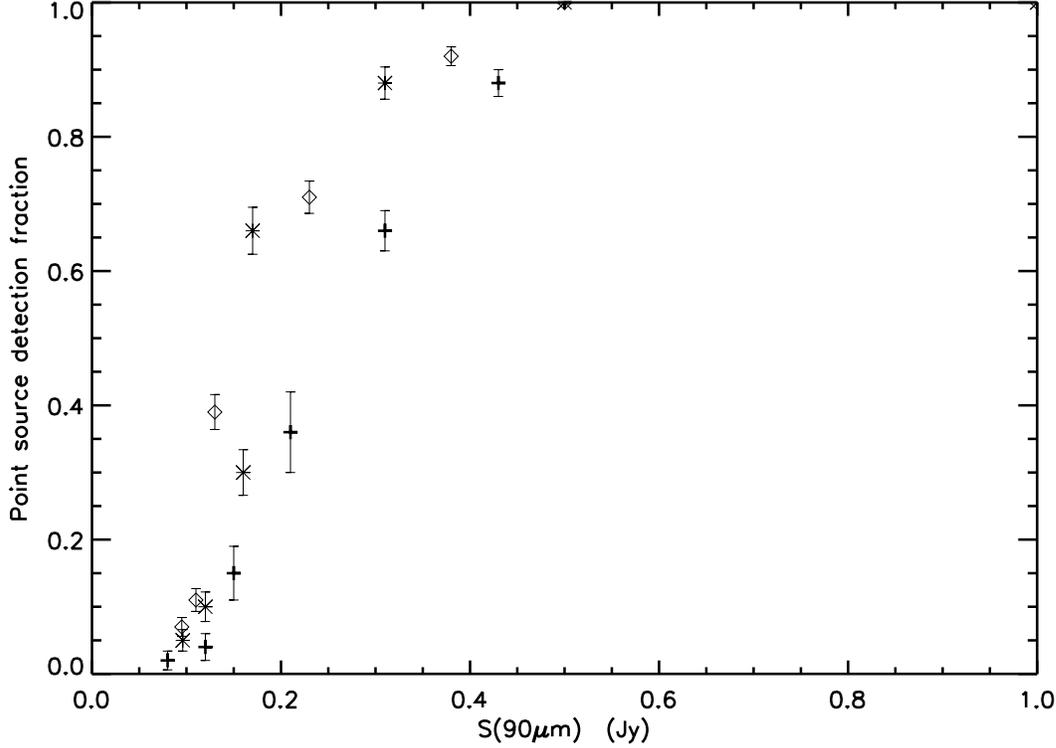,angle=0,width=15cm}
\caption{Detection fraction of point sources as a function of flux and
redundancy (diamonds=3, stars=1, crosses=0) estimated from
simulations. The uncertainties are the statistical $\sqrt{f (1-f) N}$ errors on the
 detection fraction $f$ where $N$ is the number of sources in each simulation.}
\label{fig5}
\end{figure*}

 The results of this study are plotted in Figure 5. The fraction of
 sources detected increases significantly as we move from
 no redundancy to single redundancy. Further redundancy improves the
 completeness  at faint fluxes.

\section{Source counts}

\subsection{IRAS counts}

 We have extracted 90$\mu m$ extragalactic counts at fluxes brighter
 than possible with ELAIS by using the PSCz catalogue (Saunders {\em et al.}
 2000). We applied a cut with galactic latitude ($|b|
 > 20$) which limits the areal coverage to 7.642 steradians
 (Rowan-Robinson {\em et al.} 1991) and the number of galaxies to 11405.
 We also excluded galaxies with low IRAS flags ($fqual < 3$) at
 100$\mu m$. The resulting sample consists of 8529 galaxies. The
 90$\mu m$ flux of each galaxy is esimated by linearly interpolating
 between the 60$\mu m$ and 100$\mu m$ fluxes (by design all galaxies
 in the PSCz catalogue have reliable 60$\mu m$ fluxes).

  The IRAS counts are plotted in Figure 7. The structure in the counts
  above 5Jy, which is also evident in the 60$\mu m$ counts (Hacking,
  Condon \& Houck 1987), is probably due to a local excess in the
  source density dominated by the Virgo cluster.

  The PSCz catalogue only includes galaxies with $S(60) > 0.6Jy$. IRAS
  galaxies are known to have $S(100)/S(60)$ ratios  at least as high
  as 3.5 (e.g. Rowan-Robinson \& Crawford 1989, Efstathiou {\em et
  al.}, in preparation). A $90\mu m$ sample drawn from PSCz is
  therefore bound to be incomplete below 2Jy. This incompleteness is
  most probably the reason for the flattening of the IRAS counts
  between 1-2Jy.

\subsection{ELAIS counts}

 To extract the source counts we need to determine the area over which
 the survey is sensitive to as a function of flux. To do that we use
 the results from the simulations given in Figure 5. While we have not
 simulated sources with $S>0.5Jy$ we can reasonably assume that we are
 close to 100\% complete for those fluxes as we have detected all IRAS
 sources in the fields.  We only consider regions of redundancy 3 or
 less in this paper (amounting to a total of 11deg$^2$) and therefore
 exclude the S2 area. All of the small areas of special scientific
 interest are also excluded as some are centred on known objects.  The
 resulting areal coverage as a function of flux is given in Figure 6.
 The counts are corrected for incompleteness by normalizing each
 source by the area over which each source could be found (Hacking \&
 Houck 1987, Oliver {\em et al.} 1997).  The resulting integral counts are
 given in Figure 7. Flux uncertainties of $\pm 0.15$dex are also
 indicated. The uncertainties in the number counts are the Poisson
 errors, uncertainties in the effective area have not been included.

\begin{figure*}
\epsfig{file=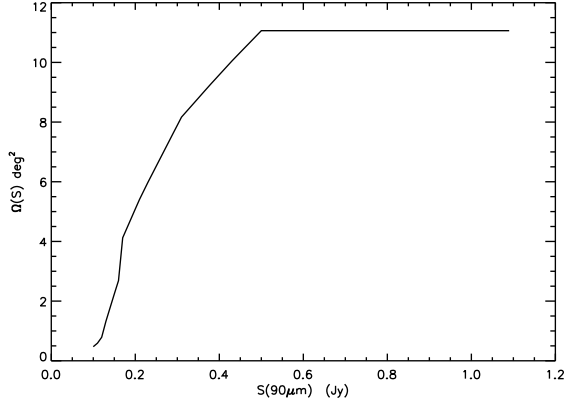,angle=0,width=8cm}
\caption{Area (in square degrees) over which  the sensitivity of the survey
 is S. The area is estimated from simulations for different values of S and 
 different degrees of redundancy.}
\label{fig6}
\end{figure*}

 The sources presented here are extracted from a reliable subset of
 the EPA sources as described in section 3. About 42\% of the sources
 brighter than 140mJy are detected in at least two neighbouring or
 overlapping pixels. About 67\% of the sources also have an
 association with a mid-infrared ELAIS source within a 1 arcminute
 radius.  Given the relatively low source density of ELAIS sources the
 probability of a chance association is only a few per cent.  We also
 note that there are good optical ids associated with the majority of
 the PHOT sources.

 To obtain an estimate of the likely contamination of our sample by
 cirrus we use the models of Gautier et al (1992). Our observing mode
 is different from either of the two cases considered in that paper as
 our reference (background) position is offset by 67'' to 390''. If we
 use their double aperture mode with this range of offset and assume
 the slope of the power spectrum of the cirrus is in the range -2.6 to
 -3.8, then the expected rms cirrus confusion noise is predicted to be
 in the range 0.2-3.1mJy for a background intensity of 1MJy/sr. The
 latter is about a factor of 2 brighter than that found in the typical
 ELAIS area (PAPER I), so it is very unlikely that any of our sources
 are due to cirrus.

 We estimate the slope of the counts (assuming they can be described by a power-law)
 by applying a  $\chi^2$ minimization procedure (the IDL routine linfit) on
 the  model

 $$\log_{10}({{dN} \over {d \log_{10} S}}) = a + b \log_{10} S  $$

 In the range 0.158-1Jy we find $a=3.71 \pm 0.13$ and $b=-1.92 \pm
 0.23$. If we use only sources found in areas with redundancy in our
 analysis, the estimated parameters are $a=3.65 \pm 0.22$ and $b=-2.11
 \pm 0.34$. We also estimate the slope of the IRAS counts in the range
 2-20Jy (where the effects of incompleteness and large scale structure
 are minimized) to be $a=3.73 \pm 0.02$ and $b=-1.48 \pm 0.03$.  The
 ELAIS counts therefore show some evidence for departure from the Euclidian
 slope.  However, because of the limited statistics of the ELAIS
 survey and the incompleteness of IRAS at around 1Jy it is difficult
 to constrain from this study the flux at which this takes place.

\section{Discussion}

\subsection{Comparison with evolutionary models}

The recent dramatic improvement in observational constraints in the
star-formation history of the Universe and therefore the evolution of
the starburst galaxy populations in the optical/UV, as well as
recently the submillimeter, has stimulated the development of
evolutionary models (Pei \& Fall 1995, Pearson \& Rowan-Robinson 1996,
Franceschini {\em et al.}  1997, Guiderdoni {\em et al.} 1998,
Rowan-Robinson 1999).  In Figures 7 and 8 we compare the observed
counts with the models of Rowan-Robinson, Guiderdoni {\em et al.} and
Franceschini {\em et al.}. The counts predicted by the models differ
by a factor of 3-4 at about 100mJy and the ELAIS counts can
potentially discriminate between them.

 The model of Guiderdoni {\em et al.} is set within the framework of
 hierarchical growth of structures according to the cold dark matter
 model, and extends earlier studies (e.g. by Cole {\em et al.} 1994) to the
 IR/submm wavelength regime. In Figure 7 we plot the prediction from
 their models A and E. The latter model incorporates a heavily
 extinguished (ULIRG) population, assumed to dominate at high redshift
 in order to account for the far-ir background.

 The starting point in the model of Rowan-Robinson (1999) is an
 assumed star formation history of the Universe, described by an
 analytic formula involving 2 free parameters $(P,Q)$. The star
 formation rate is assumed to evolve due to pure luminosity
 evolution. The parameters $P$ and $Q$ are then constrained by fits to
 the observed star formation history, the observed 60- 175 and 850$\mu
 m$ counts and the observed far-ir background. The model assumes a
 number of infrared galaxy populations which are modelled with the
 radiative transfer models of Efstathiou, Rowan-Robinson \& Siebenmorgen
 (2000). The prediction of the Rowan-Robinson model is compared with the
 observed  counts in Figure 8.

 The model of Franceschini {\em et al.} (in preparation) assumes
 that the extragalactic population is composed of two components with
 different evolution properties: (1) a non-evolving galaxy population
 in which the far-IR flux is mostly contributed by evolved stars; (2)
 a population of strongly evolving starbursts, with peak emissivity
 around redshift 0.8 and roughly constant emissivity at higher z. The
 evolution rate for population (2) is optimized to reproduce the
 mid-IR counts and redshift distributions.  The model of Franceschini
 is also plotted on Figure 8.

 While the observed counts tentatively favour the Guiderdoni model,
 the uncertainties due to flux calibration and correction for
 incompleteness do not allow us to definitely discriminate between the
 models at the present stage. A more detailed analysis of the ELAIS
 90$\mu m$ survey (already under way) should be able to discriminate
 between the models. Further discrimination of the two models should
 be possible with the redshift distributions of the detected
 galaxies. The models of Franceschini {\em et al.} and Rowan-Robinson
 predict that the vast majority of sources at $S > 100$mJy will be at
 redshifts less than 1. The Guiderdoni {\em et al.} model overpredicts
 the number of high redshift galaxies in the North Ecliptic Pole (NEP)
 survey, although the counts in the latter are known to be dominated
 by a supercluster (Ashby {\em et al.} 1996). The ELAIS survey was
 specifically designed (PAPER I) to minimise the effects of large
 scale structure.

\begin{figure*}
\epsfig{file=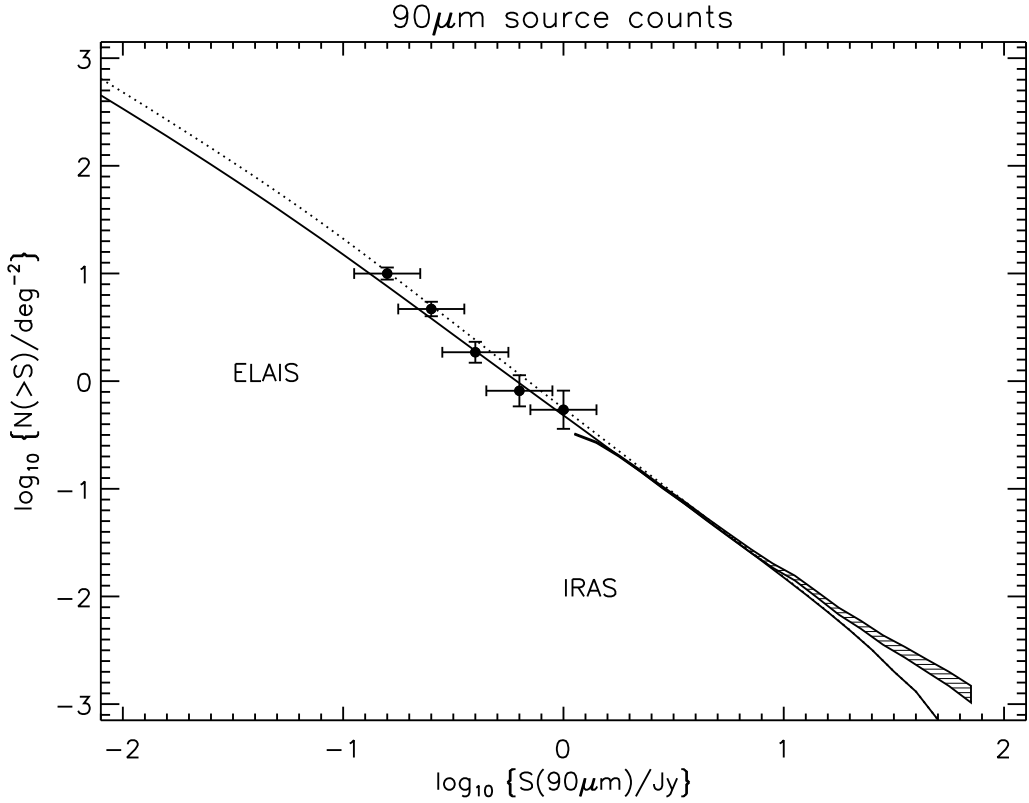,angle=90,width=15cm}
\caption{ELAIS/IRAS 90$\mu m$ source counts. The solid and dotted
 lines are the counts predicted
by the models A and E (respectively) of Guiderdoni {\em et al.} (1998).}
\label{fig7}
\end{figure*}

\begin{figure*}
\epsfig{file=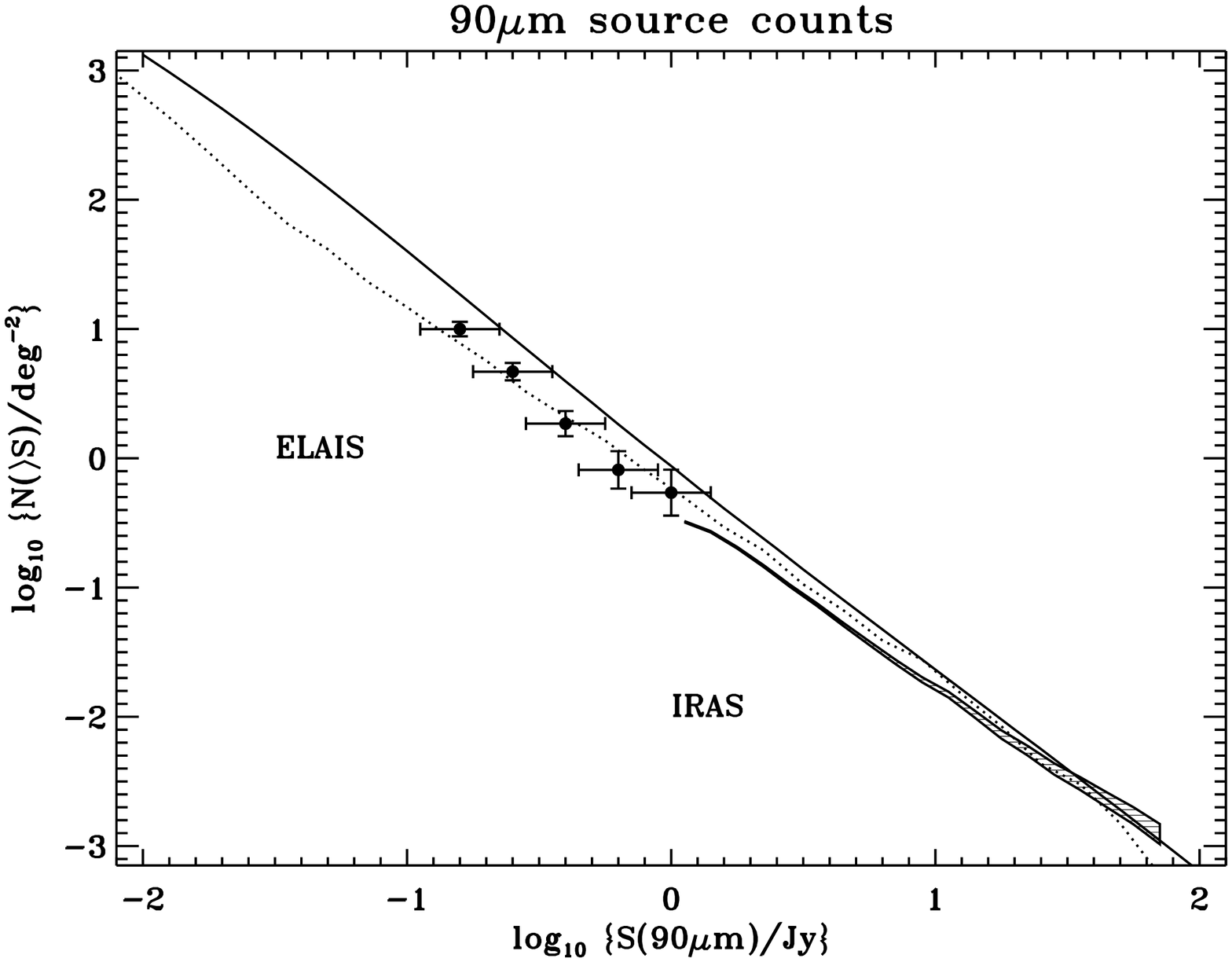,angle=90,width=15cm}
\caption{ELAIS/IRAS 90$\mu m$ source counts. The solid line gives the
counts predicted by  the model  of Rowan-Robinson (1999).
 The dotted line shows the counts predicted
by the model of Franceschini.}
\label{fig8}
\end{figure*}

\subsection{Comparison with other Infrared and submillimeter surveys}

Until recently the deep IRAS 60$\mu m$  in the NEP survey (Hacking \& Houck
1987) provided the deepest extragalactic source counts in the
infrared. Augmented with counts from shallower surveys
(e.g. Rowan-Robinson {\em et al.} 1991) they provided the basis for
studies of the evolution of the starburst galaxy population (Hacking
et al 1987, Oliver {\em et al.} 1992, 1995, Pearson \& Rowan-Robinson
1996). Unfortunately redshift measurements of the redshift sources,
which should have been able to constrain further the form of
evolution, discovered the presence of a $z=0.088$ supercluster in the NEP
field (Ashby {\em et al.} 1996).

 Oliver {\em et al.} (1997) confirmed the strong evolution seen in IRAS
 surveys with deep counts at 6.7 and 15$\mu m$ in the Hubble Deep
 Field. Counts at 15$\mu m$ well in excess of no-evolution expectations
  were also reported by Elbaz {\em et al.} (1999).
 Serjeant {\em et al.} (2000, Paper II) found the evolving high-z
 population dominates over the local counts at $S(15) < 10mJy$.

 With the advent of SCUBA, submillimeter surveys (at 850$\mu m$) which
 exploited the negative K-correction due to the spectral shape of
 starburst galaxies, detected source densities 2-3 orders of
 magnitude higher than expected from non-evolving models (Smail {\em et al.}
 1997, Hughes {\em et al.} 1998, Barger {\em et al.} 1998, Eales {\em et al.} 1999). While
 the process of identifying and determining the properties of these
 SCUBA sources is still incomplete, it appears that a number of them are
 similar to ULIRGs at $z\sim2-4$.  SCUBA surveys appear to have
 resolved 94\% and 34\% of the COBE-FIRAS background values at 850 and
 450$\mu m$ respectively (Blain {\em et al.} 1999).

 FIRBACK 175$\mu m$ counts in the Marano 1 field also appear to have
 shown an upturn in the counts (Puget {\em et al.} 1999). These counts are 3
 times higher than the models of Guiderdoni {\em et al.} and Franceschini et
 al. However, 175$\mu m$ counts in a larger area (Dole {\em et al.} 1999)
 suggest that the counts in the Marano field may have been anomalously
 high. It is also not clear whether the excess in the counts, if any,
 is due to high redshift starbursts or local cold galaxies. 
 Counts 3-10 times higher than expected from no-evolution models were also
 found by Kawara  {\em et al.} (1998) in their 175$\mu m$ survey of the Lockman Hole.
 More results are expected from FIRBACK and ELAIS surveys at this
 wavelength.  Brighter counts are also expected from the ISO
 Serendipity survey (Bogun {\em et al.} 1996, Stickel {\em et al.} 1998).

 The extensive multi-wavelength coverage and follow-up programme of
ELAIS ensures that, unlike other far-ir surveys where resolution and
sensitivity can be a serious obstacle, the 90$\mu m$ galaxies will
be thoroughly studied (e.g. Morel {\em et al.} in preparation) and will
therefore provide a firm basis for studying the evolution of
dust-enshrouded galaxies.   

A more detailed analysis of the ELAIS PHT survey
is underway and results  will be presented in subsequent papers in
this series.


\section*{Conclusions}

 We have presented $90\mu m$ source counts extracted from the
 catalogue produced by the ELAIS Preliminary Analysis and PSCz
 redshift survey. All but one of the sources brighter than 100mJy have
 been confirmed by an independent pipeline developed at MPIA,
 Heidelberg. We compare our estimated fluxes with IRAS fluxes and
 models for standard stars and estimate the uncertainty in flux
 calibration to be about 30-40\%.  

 The ELAIS counts extend the IRAS counts by an order of magnitude in
 flux.  The slope of the counts in the 0.158-1Jy range shows some evidence
 for departure from the Euclidian slope. This is consistent with the
 strong evolution seen in other infrared and submillimeter surveys.

 Within the uncertainties associated with the flux calibration of the
 survey, the counts agree with the strongly evolving models of
 Rowan-Robinson (1999), Guiderdoni {\em et al.} (1998) and
 Franceschini {\em et al.}. We expect the redshift distributions
 arising from the spectroscopic follow-up of the survey to allow us to
 discriminate between these models.

\section*{Acknowledgments}

 It is a pleasure to acknowledge Bernhard Schulz, Ulrich Klaas, Uwe
 Herbstmeier and other members of the PHOT team for very useful
 discussions.  We also thank Peter Hammersley and Martin Cohen for
 making their stellar seds available to us.  AE acknowledges support
 by PPARC.  This research has made use of the NASA/IPAC Extragalactic
 Database (NED) which is operated by the Jet Propulsion Laboratory,
 California Institute of Technology, under contract with the National
 Aeronautics and Space Administration.  ELAIS is supported by EU TMR
 Network FMRX-CT96-0068 and PPARC grant GR/K98728. We also thank an
 anonymous referee for useful comments and suggestions.

 This paper is based on observations with ISO, an ESA project with
 instruments funded by ESA member states (especially the PI countries:
 France, Germany, the Netherlands and the United Kingdom) and with
 participation of ISAS and NASA. The PHT data were  processed
 using PIA, a joint development by the ESA Astrophysics Division and
 the ISOPHOT consortium led by MPI f\"{u}r Astronomie, Heidelberg.
 Contributing Institutes are DIAS, RAL, AIP, MPIK, and MPIA.


\begin{thebibliography}{99}

\bibitem{} Ashby, M.L.N., Hacking, P., B., Houck, J.R., Soifer, B.T.,
\& Weisstein, E.W., 1996, ApJ, 456, 428.

\bibitem{}  Barger, A.J.,  Cowie, L.L., Sanders, D.B., Fulton, E.,
Taniguchi, Y., Sato, Y., Kawara, K., Okuda, H., 1998, Nature, 394, 248.

\bibitem{} Blain, A., Ivison, R., Kneib, J.-P., \& Smail, I., 1999,
 astro-ph/9908024.

\bibitem{} Bogun, S.,  {\em et al.}  1996, AA, 315, L71.

\bibitem{}  Cohen, M,  {\em et al.}, 1999, AJ, 117, 1864. 

\bibitem{} Condon, J.J., Huang, Z.-P., Yin, Q.F., Thuan, T.X.,  1991,
ApJ, 378, 65.

\bibitem{} Dole, H. {\em et al.} 1999, in 'The Universe as seen by ISO', 
    ed.P.Cox \& M.F.Kessler, p.1031.

\bibitem{} Eales, S., Lilly, S., Gear, W., Dunne, L., Bond, R.J.,
Hammer, F., Le Fevre, O., Crampton, D., {\em et al.}, 1999, ApJ, 515, 518.

\bibitem{} Efstathiou, A., \& Rowan-Robinson, M., 1995, MNRAS, 273,
649.

\bibitem{} Efstathiou, A., Rowan-Robinson, M., \& Siebenmorgen, R.,
 2000, MNRAS, 313, 734. 

\bibitem{}  Fixsen, D.J., Dwek, E., Mather, J.C., Bennett, C.L.,
Shafer, R.A., 1998, ApJ, 508, 123.

\bibitem{}  Franceschini, A., Mazzei, P., De Zotti, G., \& Danese, L., 1994, 
ApJ, 427, 140.

\bibitem{}  Gabriel, C., {\em et al.}, 1997, proceedings of the ADASS VI
Conference, eds. G.Hunt, H.E. Payne, pp.108.

\bibitem{}  Gautier, T.N., Boulanger, F., Perault, M., \& Puget, J.L., 1992, AJ, 103, 1313.

\bibitem{}  Granato, G.L., \& Danese, L., 1994, MNRAS, 268, 235.

\bibitem{}  Guiderdoni, B., Hivon, E., Bouchet, F.R., Maffei, B.,
MNRAS, 1998, 295, 877.

\bibitem{}  Hacking, P., \& Houck, J.R., 1987, ApJS, 63, 311.

\bibitem{}  Hacking, P.,  Condon, J.J., \& Houck, J.R., 1987, ApJ, 316, L15.

\bibitem{}  Hammersley, P., Jourdain de Muizon, M., Kessler, M.F.,
Bouchet, P., Joseph, R.D., Habing, J., Salama, A., \& Metcalfe, L.,
1998, AAS, 128, 207. 

\bibitem{}  Hauser, M.G., Gillett, F.C., Low, F.J., Gautier, T.N.,
Beichman, C.A., Aumann, H.H., Neugebauer, G., Baud, B., Boggess, N.,
Emerson, J.P.,  1984, ApJ, 278, 15.

\bibitem{} Hauser, M.G., {\em et al.} 1998, ApJ, 508, 25.

\bibitem{}  Hughes, D., {\em et al.}, 1998, Nature, 394, 241.

\bibitem{}  Kessler, M.,  {\em et al.},  1996, AA, 315, 27.

\bibitem{}  Kawara, K., et al., 1998, AA, 336, L9.

\bibitem{}  Klaas, U., 1994, ISOPHOT observer's manual, Version 3.1.

\bibitem{}  Kr\"{u}gel, E., Siebenmorgen, R., Zota, V., Chini, R., 1998,
AA, 331, L9.

\bibitem{}  Lagache, G., Abergel, A., Boulanger, F., Puget, J.-L.,
1998, AA, 333, 709.

\bibitem{} Laureijs, R, 1999,
{http://isowww.estec.esa.nl/users/expl-lib/PHT-top.html}

\bibitem{}  Lemke, D.,  {\em et al.},  1996, AA, 315, 64. 

\bibitem{} Madau, P., Ferguson, H.C., Dickinson, M.E., Giavalisco, M.,
Steidel, C.C., Fruchter, A., 1996, MNRAS, 283, 1388.

\bibitem{} Oliver, S., Rowan-Robinson, M., Saunders, W., 1992, MNRAS, 256,
 15p.

\bibitem{} Oliver, S.J.,  {\em et al.}, 1995, in 'Wide Field spectroscopy and the 
 distant universe', World scientific, ed.S.J.Maddox \& A.Aragon-Salamanca, 
 p.274.

\bibitem{} Oliver, S.J.,  {\em et al.}, 2000, in press. astro-ph/0003263 PAPER I

\bibitem{} Pearson, C., \& Rowan-Robinson, M., 1996, MNRAS, 283, 174.

\bibitem{} Pei, Y.C. \& Fall, S.M., 1995, ApJ, 454, 69.

\bibitem{} Pier, E, \& Krolik, J., 1992, ApJ, 399, L23. 

\bibitem{}  Puget, J.-L.,  Abergel, A., Bernard, J.-P., Boulanger, F.,
Burton, W.B., Desert, D.-X., Hartmann, D., 1996, AA, 308, L5.

\bibitem{}  Puget, J.-L.,  {\em et al.}, 1999, AA, 345, 29

\bibitem{} Reach, W.T., Abergel, A., Boulanger, F.,  {\em et al.}, 1996, AA,
315, L381.

\bibitem{} Rowan-Robinson, M., \& Crawford, J., 1989, MNRAS, 238, 523.

\bibitem{} Rowan-Robinson, M., Saunders, W., Lawrence, A., \& Leech,
K., 1991, MNRAS, 253, 485.

\bibitem{} Rowan-Robinson, M., 1999, astroph/9906308

\bibitem{} Rowan-Robinson, M., \& Efstathiou, A., 1993, MNRAS, 263, 675.

\bibitem{} Saunders, W., {\em et al.}, 2000, MNRAS, submitted, astro-ph/0001117.

\bibitem{}  Schlegel, D., Finkbeiner, D., \& Davis, M., 1998, ApJ,
500, 525.


\bibitem{}  Schulz, B.,  {\em et al.} 1999, {http://isowww.estec.esa.nl/users/expl-lib/PHT-top.html}

\bibitem{} Serjeant, S.,  {\em et al.},  1999, MNRAS, in press astro-ph/0003198 (PAPER II)

\bibitem{} Smail, I., Ivison, R.J., \& Blain, A.W., 1997, ApJ, 490, L5.

\bibitem{} Steidel, C.C., Giavalisco, M., Pettini, M., Dickinson, M.,
\& Adelberger, K.,L., 1996, ApJ, 462, L17.1996

\bibitem{} Stickel, M., Lemke, D., Klaas, U., Toth, L.V., Herbstmeier,
UU., Richter, G., Assendorp, R., Laureijs, R., Kessler, M.F.,
Burgdorf, M., Beichman, C.A., Rowan-Robinson, M., Efstathiou, A.,
1998, AA, 336, 116.
   
\bibitem{}
   
\bibitem{}
   
\end{thebibliography}
\end{document}